\documentclass[runningheads]{llncs}
\usepackage{graphicx}
\usepackage{subcaption}
\usepackage{amsmath}
\usepackage{amsfonts}
\renewcommand*\vec[1]{\boldsymbol{\mathbf{#1}}}
\DeclareMathOperator{\argmax}{arg\,max}

\begin{document}
\title{Reconstructing MRI parameters using a noncentral chi noise model}
\author{Klara Ba\'s\inst{1,2} \and
Christian Lambert\inst{1} \and
John Ashburner\inst{1}\orcidID{0000-0001-7605-2518}}
\authorrunning{K. Ba\'s, et al.}
%
\institute{
\emph{Wellcome Centre for Human Neuroimaging, UCL, UK}\\
\and
\emph{Department of Medical Physics \& Biomedical Engineering, UCL, UK}\\
%
}
\maketitle              
\begin{abstract}
Quantitative magnetic resonance imaging (qMRI) allows images to be compared across sites and time points, which is particularly important for assessing long-term conditions or for longitudinal studies. The multiparametric mapping (MPM) protocol is used to acquire images with conventional clinical contrasts, namely PD-, T1-, and MT-weighted volumes. Through multi-echo acquisition for each contrast and variations in flip angles between PD- and T1-weighted contrasts, parameter maps, such as proton density (PD), longitudinal relaxation rate (R1), apparent transverse relaxation rate (R2$^*$), and magnetization transfer saturation (MT$_{sat}$), can be estimated. Various algorithms have been employed to estimate these parameters from the acquired volumes. This paper extends an existing maximum a posteriori approach, which uses joint total variation regularization, by transitioning from a Gaussian noise approximation to a more physically plausible model that assumes noncentral chi-distributed noise.

\end{abstract}
\section{Introduction}
The aim of quantitative MRI (qMRI) is to obtain images with voxel intensities that represent the underlying tissue properties and which have physical units. This differs from conventional MRI used in clinical practice, where voxel values are relative and the scanner sequences are chosen to maximise the contrast among tissues of interest. Voxel brightness in sequences known as, for example, T1-weighted are in fact affected by a number of factors~\cite{Yokoo2010}.
The aim of qMRI is to compute voxel values that characterise the underlying tissue. This is especially important for studies of long-term conditions, such as Parkinson's disease, when patients are scanned at different time points and often on different scanners. The introduction of magnetic resonance with quantitative values should allow greater homogeneity across data collected in such studies.

One of the proposed scanning protocols is multi-parameter mapping (MPM)~\cite{Weiskopf2008}. This protocol consists of three 3D FLASH multi-echo acquisitions, which could conventionally be referred to as ``PD-'', ``T1-'' and ``MT-weighted''. These three multi-echo runs are used to estimate four parameter maps: proton density (PD), longitudinal relaxation (R1 = 1/T1), magnetization transfer saturation  (MT$_{sat}$), and apparent transverse relaxation (R2$^*$ = 1/T2$^*$).

Noise in MR images is often modelled by Gaussian or Rice distributions, which may suffice for some purposes. However, when accounting for the image reconstruction process one needs also to propagate the initial Gaussian electronic noise. The noise on the individual receivers is indeed Gaussian, but when real and imaginary (two orthogonal coils) are combined the noise on the magnitude image is Rice distributed. In brain imaging, a multi-coil scanning setup is commonly used, which involves images that are reconstructed by combining signals from multiple coils. There are different reconstruction protocols, such as GRAPPA~\cite{Griswold2002} or SENSE~\cite{Pruessmann1999}. Reconstruction from multiple coils means that noise itself is propagated, leading to a change in the noise distribution. In the case of GRAPPA, reconstruction noise can be better approximated by a noncentral chi (nc-$\chi$) distribution~\cite{AjaFernandez2014}. This paper describes extending the existing \emph{NITorch}\footnote{\url{https://github.com/balbasty/nitorch}} maximum a posteriori model fitting algorithm for estimating MPM parameters~\cite{Balbastre2021} so that it can use the assumption that MRI noise is nc-$\chi$ distributed.


\section{Theory}
For a nc-$\chi$ distribution, the probability density function (pdf) of a statistic $s$ is given as a function of the number of independent random variables $\nu$ and another statistic $\lambda$:
\begin{equation}
    p(s | \nu, \lambda) = \frac{e^{\frac{-(s^2+\lambda^2
)}{2}}s^{\nu}\lambda}{(\lambda s)^{\frac{\nu}{2}}}I_{\frac{\nu}{2}-1}(\lambda s).
\end{equation}
In the above equation, $I_{K}(\cdot)$ denotes the modified Bessel function of the first kind, of order $K$.
The statistics in the equation are given by:
\begin{equation}
    s = \sqrt{\sum_{i=1}^{\nu}\left(\frac{z_i}{\sigma_i}\right)^2}
    \text{ and }
    \lambda = \sqrt{\sum_{i=1}^\nu\left(\frac{\mu_i}{\sigma_i}\right)^2}
    \text{,}
    \label{eq:ncchi_ss}
\end{equation}
where $z_i$ are independent, normally distributed random variables with means $\mu_i$ and variances $\sigma^2_i$.

For our proposed work, we introduce the notation $\mu = \sqrt{\sum_i \mu_i^2}$ and all variances ($\sigma^2_i$) are assumed equal. This allows the probability density to be re-expressed by substituting $s = x/\sigma$ and $\lambda = \mu/\sigma$, with an additional division by $\sigma$ (motivated through integration by substitution) to ensure the probability distribution integrates to 1:
\begin{equation}
    p(x | \mu, \nu, \sigma^2) = \frac{e^{\frac{-(x^2+\mu^2
)}{2\sigma^2}}x^{\nu}\mu}{\sigma^2(\mu x)^{\frac{\nu}{2}}}I_{\frac{\nu}{2}-1}\!\left(\frac{\mu x}{\sigma^2}\right). \label{eq:ncChi_pdf}
\end{equation}
For $\nu=2$ this reduces to a Rice distribution.

This work also uses the chi distribution (where the means $\mu_i$ are all zero), which has a pdf of:
\begin{equation}
    p(s|\nu) = \frac{s^{\nu-1}e^{-\frac{s^2}{2}}}{2^{\frac{\nu}{2}}\Gamma\left(\frac{\nu}{2}\right)}, \label{eq:Chi_ll}
\end{equation}
where $s$ is given by eq.~\ref{eq:ncchi_ss}. Similarly to the derivation of eq.~\ref{eq:ncChi_pdf}, it can be re-written in the following form, which is a generalisation of the Rayleigh ($\nu=2$) and Maxwell-Boltzmann ($\nu=3$) distributions:
\begin{equation}
    p(x|\nu,\sigma) = \frac{2^{1-\nu/2}x^{\nu-1}e^{-\frac{x^2}{2\sigma^2}}}{\sigma^{\nu}\Gamma\left(\frac{\nu}{2}\right)}.
    \label{eq:Chi_pdf}
\end{equation}

\subsection{Noise estimation}
In our proposed work, parameter map estimation uses pre-computed noise estimates, which are found by fitting a mixture of two distributions to the image intensities. One distribution corresponds to the image noise in the air outside the head, while the other represents all the tissue intensities~\cite{Ashburner2013}. Under the assumption that noise is uniform across the entire field of view, this allows a single noise estimate to be made from the air. Previous work fitted a mixture of two Gaussians or two Ricians, but we replace this with a mixture of two chi distributions. With purely thermal noise, one would expect the degrees of freedom $\nu$ to be equal to twice the number of channels in the head coil. However, other sources of noise are present, and we find that estimating $\nu$ from the data allows a better fit to be obtained.  The estimation of $\nu$ and $\sigma^2$ is done using the Expectation-Maximisation (EM) algorithm.

There is almost no MRI signal from the air outside the head, so the noise distribution there can be approximated by a chi distribution. This is a special case of the nc-$\chi$ distribution with $\lambda = 0$. 
The MR images are reconstructed from signal from a number of coils (the number of coils proportional to the parameter $\nu$). The means of the signal received by each coil should be $0$ due to the simplifying assumption that there is no source of signal outside of the imaged subject. We need to derive parameter updates specific to the chi distribution, and start by deriving the log-likelihood.

The general form of log-likelihood assuming $K$ mixture components is:
\begin{equation}
\ln p(\vec{x} | \vec{\theta}) = \sum_{n=1}^N \ln\left(\sum_{k=1}^{K} p(x_n | z_{kn}, \vec{\theta}) p(z_{kn} | \vec{\theta})\right),
\end{equation}
where univariate data is denoted by $\vec{x} \in \mathbb{R}^{1 \times N}$.
Latent variables are assumed to be represented by a one-hot encoding with $\vec{Z} \in \mathbb{R}^{K \times N}$. The log-likelihood of our chi distribution (from eq.~\ref{eq:Chi_pdf}) is:
\begin{equation}
    \ln{p(x_n|\nu_k,\sigma_k)} = \left(1-\frac{\nu_k}{2}\right)\ln{2} + (\nu_k-1)\ln{x_n}-\frac{x_n^2}{2\sigma_k^2}-\nu_k\ln{\sigma_k} - \ln{\Gamma\left(\frac{\nu_k}{2}\right)}.
\end{equation}
In the E-step we evaluate the responsibilities using the current parameter estimates $\vec{\theta} = \{\boldsymbol{\nu}, \boldsymbol{\sigma}, \boldsymbol{\pi} \}$:
\begin{align}
r_{kn}  = p (z_{kn} | x_n, \nu_k, \sigma_k, \pi_k) 
 = \frac{\pi_k \chi (x_n | \nu_k, \sigma_k)}{\sum_{j=1}^{K} \pi_j \chi(x_n | \nu_j, \sigma_j)},
\end{align}
where $\vec{\pi} \in \mathbb{R}_+^K$ denotes mixing proportions, with $\sum_{k=1}^K \pi_k = 1$. In our case of a mixture describing the subject and image background, $K=2$. Keeping the responsibilities $r_{kn}$ constant in the M-step, we evaluate:
\begin{equation}
\vec{\theta} = \argmax_{\vec{\theta}}\sum_{n=1}^{N}\sum_{k=1}^{K} r_{kn} \ln p(x_n, z_{kn} | \vec{\theta}). \label{param_upd}
\end{equation}
Mixing proportion updates have the same form as for a mixture of Gaussian distributions. This is because the derivative with regard to the mixing proportions is not dependent on the distribution. This is given by:
\begin{equation}
\pi_k = \frac{\sum_{n=1}^{N}r_{kn}}{\sum_{j=1}^{K}\sum_{n=1}^{N} r_{jn}}.
\end{equation}
Updates for the parameters $\nu_k$ and $\sigma_k$ have no closed form solutions, so they are done using an additional iterative procedure. Using eq.~\ref{eq:Chi_ll} we have:
\begin{equation}
    \frac{\partial}{\partial\sigma_k} \ln p(x_n|\nu_k, \sigma_k) = -\frac{x_n^2}{\sigma_k^{3}} + \frac{\nu_k}{\sigma_k}. \label{der_sig}
\end{equation}
This is dependent on the parameter $\nu_k$, which is why iterative updates are needed. Using q.~\ref{der_sig} to compute the partial derivative of eq.~\ref{param_upd}, and setting the result to zero gives:
\begin{align}
    \frac{\partial}{\partial\sigma_k} \sum_{n=1}^N r_{kn} \ln p(x_n|\nu_k, \sigma_k) = 0 \cr
    \sum_{n=1}^N r_{kn}\left(-\frac{x_n^2}{\sigma_k^{3}} + \frac{{\nu}_k}{\sigma_k} \right) = 0 \cr
    \frac{\nu_k \sum_{n=1}^N r_{kn}}{\sigma_k} - \frac{\sum_{n=1}^N r_{kn} x_n^2}{\sigma_k^{3}} = 0 \cr \sigma_k^2 = \frac{\sum_{n=1}^N r_{kn} x_n^2}{\nu_k \sum_{n=1}^N r_{kn}} .
\end{align}
Similarly we require the derivative with respect to the parameter $\nu_k$ of eq.~\ref{eq:Chi_ll}, which is:
\begin{equation}
    \frac{\partial}{\partial\nu_k} \ln p(x_n|\nu_k, \sigma_k)
    = \frac{\ln{2}}{2} - \ln{x_n} +  \ln{\sigma_k} + \frac{1}{2}\psi\left(\frac{\nu_k}{2}\right),
\end{equation}
where $\psi(\cdot)$ is the digamma function, which is the first derivative of the logarithm of the gamma function. This is a special case of the polygamma function, which has order zero. The more general polygamma function of order $n$ (written as $\psi^{(n)}(\cdot)$) gives the $(n-1)$th derivative of the log of the gamma function. Using this, the second derivative of eq.~\ref{eq:Chi_ll} is given by:
\begin{equation}
    \frac{\partial^2}{\partial\nu_k^2} \ln p(x_n|\nu_k, \sigma_k) = -\frac{\psi^{(1)}\left(\frac{\nu_k}{2} \right)}{4},
\end{equation}
which uses the first order polygamma function. We therefore have gradient $g$ and Hessian $h$:
\begin{align}
    g = -\frac{\partial}{\partial\nu_k} \sum_{n=1}^N r_{kn} \ln p(x_n|\nu_k, {\sigma}_k) \cr
    = \sum_{n=1}^N r_{kn} \left(\ln{x_n} -\frac{\ln{2}}{2} -  \ln{\sigma_k} - \frac{1}{2}\psi\left(\frac{\nu_k}{2}\right)\right)\cr
    = \sum_{n=1}^N r_{kn} \ln{x_n} - \left(\frac{\ln{2}}{2} +  \ln{{\sigma}_k} + \frac{1}{2}\psi\left(\frac{\nu_k}{2}\right)\right)\sum_{n=1}^N r_{kn} \label{Eq:grad_nu}
\end{align}
\begin{equation}
    h = -\frac{\partial^2}{\partial\nu_k^2} \sum_{n=1}^N r_{kn} \ln p(x_n|\nu_k, {\sigma}_k)
    = \frac{\psi^{(1)}\left(\frac{\nu_k}{2} \right)}{4} \sum_{n=1}^N r_{kn}. \label{Eq:hess_nu}
\end{equation}
The Newton update is:
\begin{equation}
\nu^{\text{new}}_k = \nu^{\text{old}}_k - h^{-1}g.
\end{equation}
Updates are alternatingly applied to parameters $\nu$ and $\sigma$ (for each $k$) until convergence is reached. Only after that we return to E-step of the EM algorithm.

\subsection{Parameter map estimation}
In \emph{NITorch}~\cite{Balbastre2021}, parameter map estimation is implemented as a \emph{maximum a posteriori} (MAP) optimisation. A prior distribution serves as regularisation, and by default is a joint total variation (JTV).
The Gaussian noise approximation used in~\cite{Balbastre2021} leads to a likelihood term that involves minimising the residual sum of squares. In our proposed method, the noise model is assumed to be nc-$\chi$, so different optimisation equations are required.

The log-likelihood for our nc-$\chi$ distribution (eq.~\ref{eq:ncChi_pdf}) is:
\begin{equation}
    \ln p(x | \mu, \nu, \sigma^2) = \frac{-(x^2+\mu^2)}{2\sigma^2}+\ln\left(x^{\frac{\nu}{2}}\mu^{1-\frac{\nu}{2}}\right)-\ln(\sigma^2)+\ln\left(I_{\frac{\nu}{2}-1}\left(\frac{\mu x}{\sigma^2}\right)\right).
\end{equation}

At each voxel, $\mu$ is obtained as a function of the qMRI forward model~\cite{Balbastre2021} using parameters $\vec{\theta}$, where $\theta_1$ denotes the longitudinal relaxation rate (R1), $\theta_2$ denotes the transverse relaxation rate (R2$^*$), $\theta_3$ denotes the proton density and $\theta_4$ denotes the magnetisation transfer saturation.
Acquisition settings over the $N$ scans are denoted by $\vec{S}$, where for scan $n$, $s_{1n}$ denotes the repetition time (TR), $s_{2n}$ is the echo time (TE), $s_{3n}$ is the flip angle, $s_{4n}$ indicates whether or not there was a magnetisation transfer pulse and $s_{5n}$ is a second repeat time for the MT-weighted scans.
This gives:
\begin{equation}
\mu(\vec{\theta}; \vec{s}) = 
\begin{cases}
\theta_3  \sin s_3 \frac{1 + (\theta_4-1)\exp(-(s_1+s_5) \theta_1) -\theta_4 \exp(-s_5 \theta_1)}{1+(\theta_4-1)\cos s_3 \exp(-(s_1+s_5)\theta_1)} \exp(-s_2 \theta_2) & \text{ if } s_4=1 \\
\theta_3 \sin s_3 \frac{1-\exp(s_1 \theta_1)}{1-\cos s_3 \exp(-s_1 \theta_1)} \exp(-s_2 \theta_2) & \text{ if } s_4=0
\end{cases}.
\end{equation}

Estimating the parameter maps involves a separate log-likelihood at each voxel.
For a single voxel, we can then define the likelihood part of the objective function to be minimised by:
\begin{equation}
    \mathcal{E}(\vec{\theta}) = - \sum_{n=1}^N \ln p(x_n | \mu(\vec{\theta}; \vec{s}_{n}), \nu, \sigma^2).
\end{equation}

As mentioned previously, the parameters are optimised using Newton updates, which requires the gradient ($\vec{g}(\vec{\theta})$) and a suitable approximation of the Hessian ($\vec{H}(\vec{\theta})$) of the above objective function for each voxel.
Gradients are given by:
\begin{align}
    g_i(\vec{\theta}) = \frac{\partial \mathcal{E}}{\partial \theta_i} = &
    \sum_{n=1}^N \left( \frac{\mu(\vec{\theta}; \vec{s}_n) - \xi_n x_n}{\sigma^2} \right)
    \frac{\partial \mu(\vec{\theta}; \vec{s}_n)}{\partial \theta_i},\cr
    \text{where }
\xi_n = & \frac{I_{\frac{\nu}{2}}\left( \frac{x_n\mu(\vec{\theta}; \vec{s}_n)}{\sigma^2} \right)}{I_{\frac{\nu}{2}-1}\left( \frac{x_n\mu(\vec{\theta}; \vec{s}_n)}{\sigma^2} \right)}.
\label{eq:gradients}
\end{align}
For robust computations, a more positive definite approximation to the Hessian (in the Loewner ordering sense) is used in the algorithm:
\begin{equation}
h_{ij}(\vec{\theta}) = \frac{1}{\sigma^2} \sum_{n=1}^N \left(
\frac{\partial \mu(\vec{\theta}; \vec{s}_n)}{\partial \theta_i}
\frac{\partial \mu(\vec{\theta}; \vec{s}_n)}{\partial \theta_j}
+ \left(\mu(\vec{\theta}; \vec{s}_n) - \xi_n x_n \right)
\frac{\partial^2 \mu(\vec{\theta}; \vec{s}_n)}{\partial \theta_i \theta_j}
\right). \label{eq:Hessian}
\end{equation}

These gradients (eq.~\ref{eq:gradients}) and Hessians (eq.~\ref{eq:Hessian}) are then plugged into the \emph{NITorch} code~\cite{Balbastre2021} in order to reconstruct parameter maps.


\section{Validation Methods}
Cross-validation was used to compare the proposed nc-$\chi$ model against the existing Gaussian model. Firstly, parameter maps were estimated from the MPM acquisitions with one of the echoes removed. These parameter maps were used, together with the forward signal model, to predict the voxel intensities of the left-out echo. This predicted echo was then compared with the left-out echo and the mean squared error (MSE) calculated. This was repeated six times for each subject, each time with a different echo removed. The MSE for the nc-$\chi$ and Gaussian models was then compared.
While calculating the MSE for the Gaussian model is straightforward ($\mathbb{E}[x] = \mu$), the nc-$\chi$ model required calculating the expected signal, which is given by:
\begin{equation}
\mathbb{E}[x] = \sigma \sqrt{\frac{\pi}{2}}{L}_{1/2}^{(\nu/2-1)}\left(\frac{-\mu^2}{2\sigma^2}\right),
\end{equation}
where $L$ is a generalised Laguerre polynomial.

\subsection{Dataset and preprocessing}
The method was evaluated using data from 19 subjects of an MPM dataset collected as a part of the \emph{qMAP-PD}
study, which is currently taking place at the Wellcome Centre for Human Neuroimaging.
\emph{qMAP-PD} has data from participants with a Parkinson’s disease diagnosis (no more than two years from diagnosis), participants diagnosed with REM sleep behaviour disorder, and healthy controls. Data acquisition followed the MPM protocol~\cite{Weiskopf2008} with 0.8 mm isotropic resolution on a 3T Siemens Prisma with a 64 channel head coil.

All volumes from the MPM dataset were registered to the first PD-weighted echo for each subject by maximising normalised mutual information using the \emph{SPM} software's coregister function.


\section{Results and Discussion}
\begin{figure}
\begin{subfigure}{0.32\textwidth}
\includegraphics[width=\textwidth]{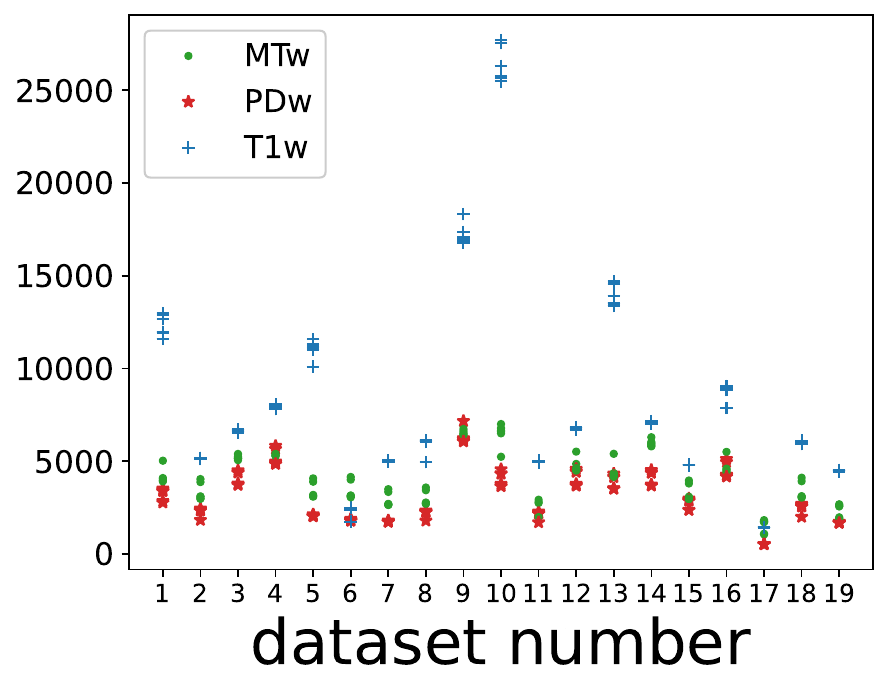}
\caption{Rice variance}
\label{figg}
\end{subfigure}
\begin{subfigure}{0.32\textwidth}
\includegraphics[width=\textwidth]{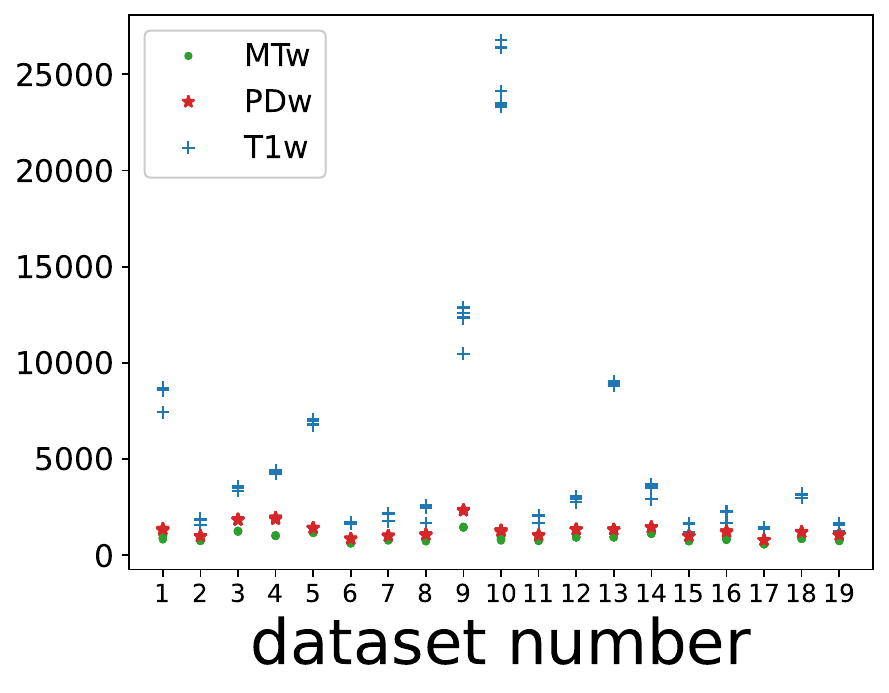}
\caption{nc-$\chi$ variance}
\label{figch}
\end{subfigure}
\begin{subfigure}{0.32\textwidth}
\includegraphics[width=\textwidth]{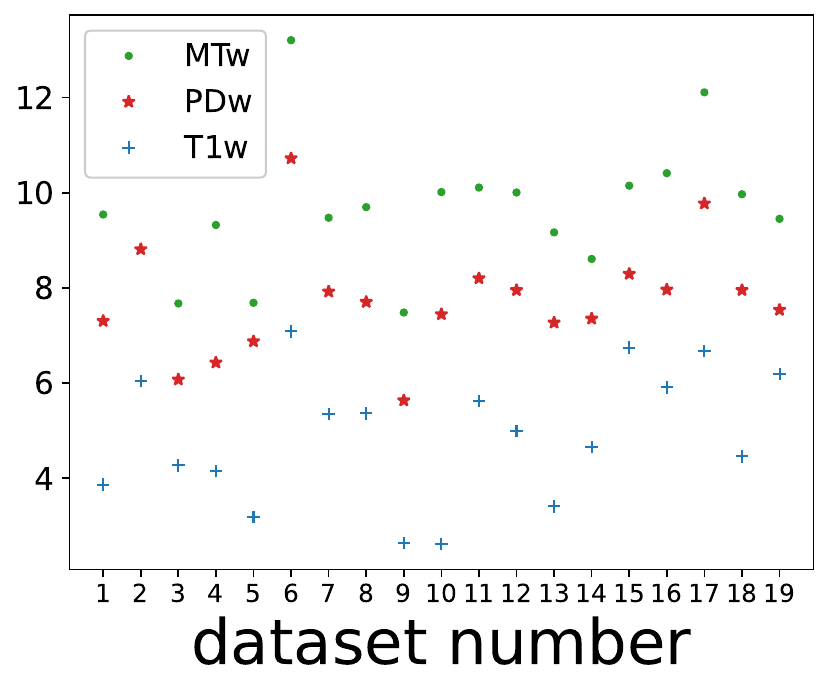}
\caption{Degrees of freedom}
\label{figd}
\end{subfigure}
\caption{Noise parameters estimated using the Rice and nc-$\chi$ noise models.}
\label{fig1}
\end{figure}

Figure~\ref{fig1} shows the image noise variance estimated using the Rice (\ref{figg}) and chi (\ref{figch}) mixture models. The variance estimates differ between the two, with a slightly higher estimate from the Rice model. Both mixture models estimate the highest variance for the T1-weighted scans, and generate what appear to be outlier variance estimates for a number of subjects.

\begin{figure}
\includegraphics[width=\textwidth]{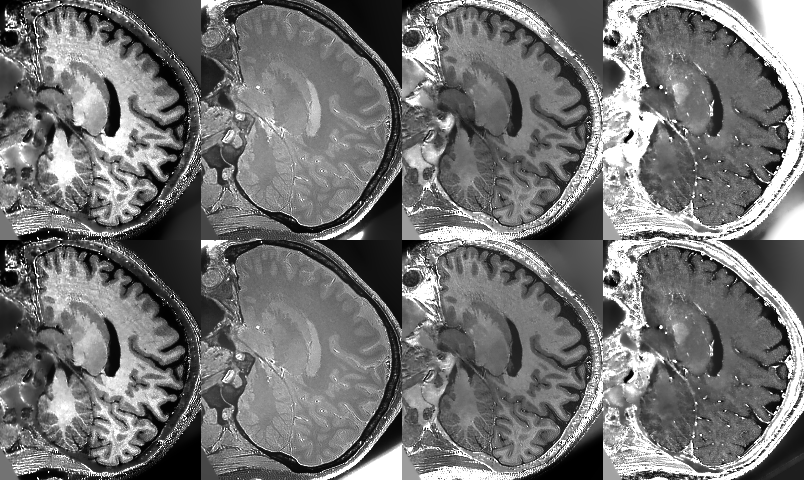}
\caption{Parameter maps. Top row: parameter maps estimated using the nc-$\chi$ model. Bottom row: parameter maps estimated with the Gaussian model. From left to right: MT, PD, R$_1$, R$_2^*$.} \label{fig2}
\end{figure}

\begin{figure}
\includegraphics[width=\textwidth]{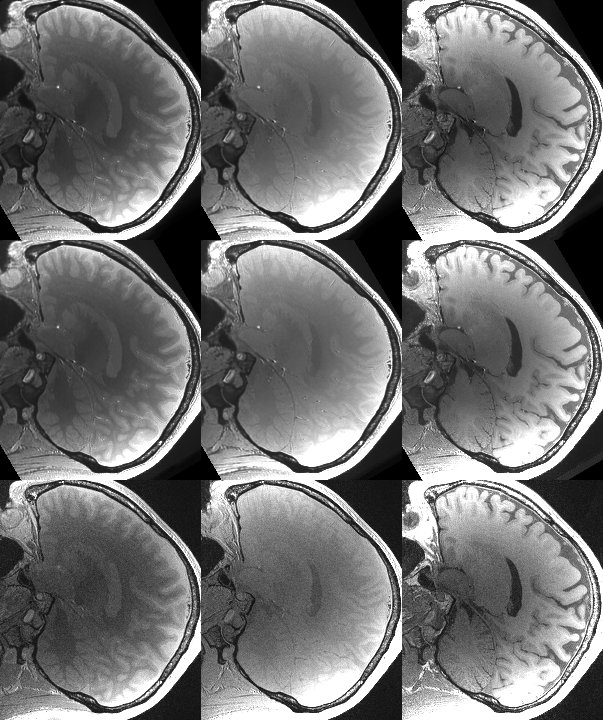}
\caption{Left out echoes (bottom row) predicted using the nc-$\chi$ model (top row), and the Gaussian model (middle row). From left to right: MTw, PDw, T1w contrasts.} \label{fig4}
\end{figure}

\begin{figure}
\begin{subfigure}{0.32\textwidth}
\includegraphics[width=\textwidth]{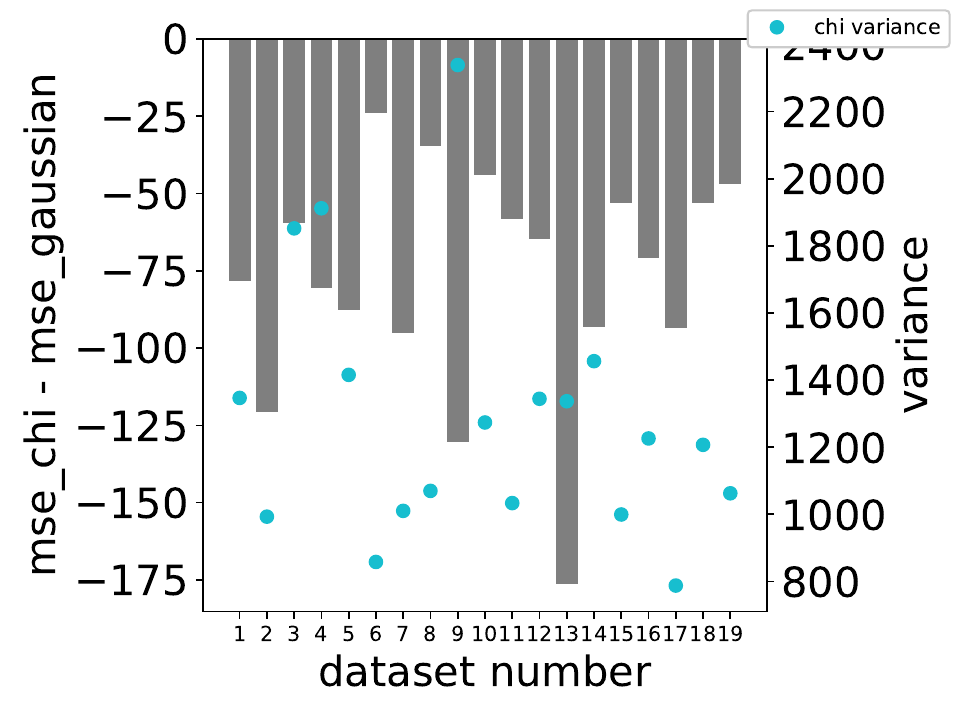}
\caption{PDw}
\end{subfigure}
\begin{subfigure}{0.32\textwidth}
\includegraphics[width=\textwidth]{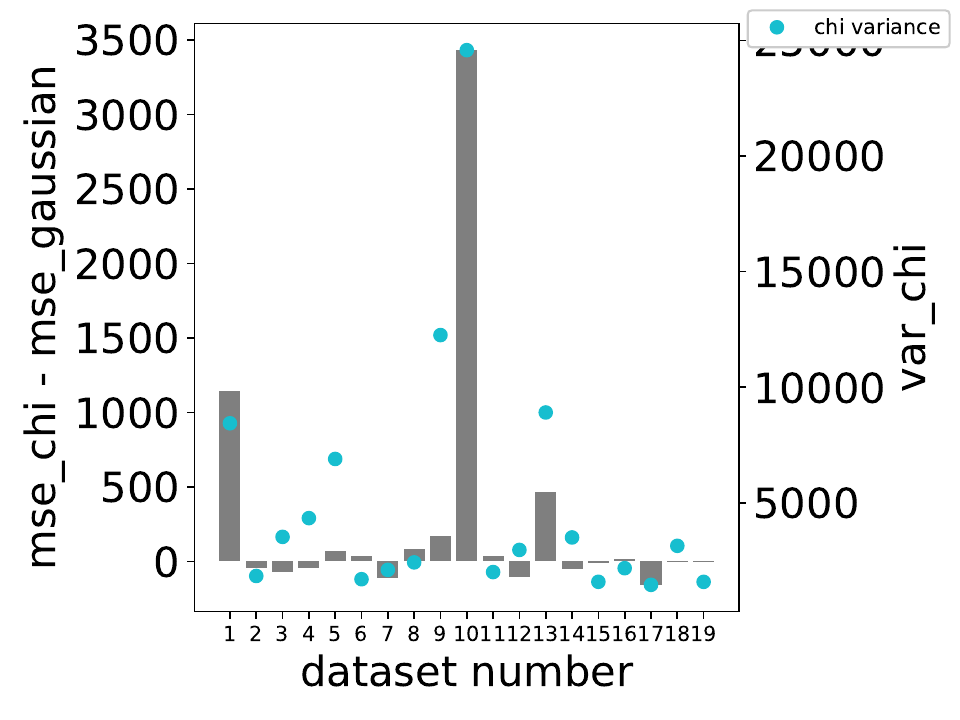}
\caption{T1w}
\end{subfigure}
\begin{subfigure}{0.32\textwidth}
\includegraphics[width=\textwidth]{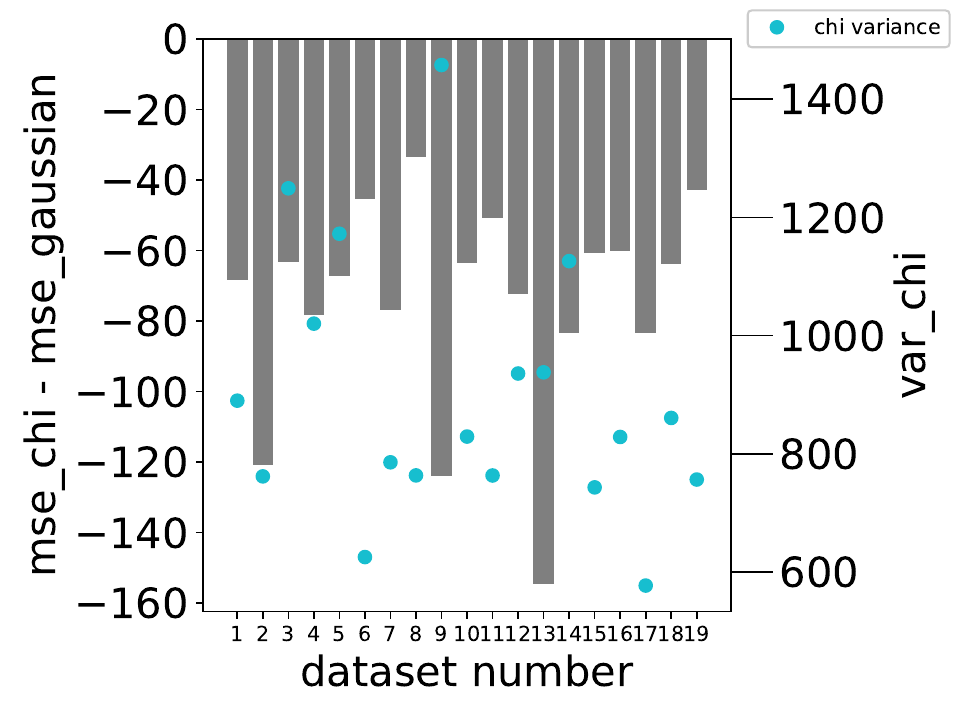}
\caption{MTw}
\end{subfigure}
\caption{Difference between MSE calculated using the nc-$\chi$ and Gaussian models. For each left-out echo, the MSE between the observed and predicted expectation of the echo was calculated. Results are averaged over all echoes. The blue dots shows the MSE estimated for the nc-$\chi$ model.}
\label{fig5}
\end{figure}

To evaluate the parameter map estimation models, we use noise estimates from the chi mixture model for both Gaussian and nc-$\chi$ parameter map estimations.
Figure~\ref{fig2} shows parameter maps estimated with the Gaussian and nc-$\chi$ models, while Figure~\ref{fig4} shows an example of a predicted echo next to the observed data.
Fig.~\ref{fig5} shows the difference between the MSE calculated for the nc-$\chi$ model and the MSE calculated for the Gaussian model. While for PD-weighted and MT-weighted scans the predictions calculated using the nc-$\chi$ model give smaller MSE, the MSE for the T1-weighted scans varied among subjects. The poorer performance of the nc-$\chi$ model seems to occur only for subjects for which extremely high noise variance was estimated from the T1-weighted scans.

The noise parameters estimated with the nc-$\chi$ model are a variance, $\sigma^2$, and the number of degrees of freedom, $\nu$. The estimated $\nu$ is typically smaller than twice the number of MRI coils (Figure~\ref{figd}), and differs among acquisitions.~\cite{AjaFernandez2014} mentions that the reconstruction increases the correlation between coils, so the estimated number of coils decreases. Finally, the noise parameters are estimated in the background, where sources other than thermal noise are present. The degrees of freedom parameter could be replaced with the "effective number of coils" consisting of the actual number of coils scaled by a parameter.


\section{Conclusion}
The nc-$\chi$ model gave better predictions for the PD and MT-weighted echoes, but the results were more variable for the T1-weighted echoes.
This appears to be because the noise estimation from the T1-weighted scans may have failed for some reason, which we will investigate further.
Another limitation is that with multi-coil imaging we would expect the noise to be non-stationary. However, this paper focused on comparing the current stationary Gaussian/Rice assumption with a stationary nc-$\chi$ noise approximation. In conclusion, the usefulness of the approximation may depends on other image aspects. The results suggest that when the noise variance is correctly specified, the nc-$\chi$ noise assumption can be beneficial.

\subsection{Acknowledgement}
KB is supported by the EPSRC-funded UCL Centre for Doctoral Training in
Intelligent, Integrated Imaging in Healthcare (i4health) (EP/S021930/1). This research was supported by NVIDIA and utilized A30 Tensor Core GPU. The Wellcome Centre for Human Neuroimaging was supported by a Wellcome Trust Centre award 539208.
This preprint has not undergone any post-submission improvements or corrections. The Version of Record of this contribution is published in Medical Image Understanding and Analysis. MIUA 2024. Lecture Notes in Computer Science, vol 14860. Springer, Cham., and is available online at https://doi.org/10.1007/978-3-031-66958-3\_13”. 

%
%
\bibliographystyle{splncs04}

\end{document}